\documentclass[12pt]{iopart}
\usepackage{iopams,setstack}
\usepackage{amsMatrix}
\usepackage{graphicx,ifthen}

\DeclareFontFamily{U}{msb}{}%
\DeclareFontShape{U}{msb}{m}{n}{<-6>msbm5<6-8>msbm7<8->msbm10}{}%
\DeclareSymbolFont{AMSb}{U}{msb}{m}{n}
\DeclareSymbolFontAlphabet{\mathbb}{AMSb}
\newcommand\R{\mathbb{R}}

\newtheorem{definition}{Definition}
\newtheorem{theorem}{Theorem}
\newtheorem{property}[definition]{Property}
\newtheorem{lemma}{Lemma}

\newcommand\w{\wedge}
\newcommand\inn{{\cdot}}
\let\lslash\l
\renewcommand\l{\ifmmode\lambda\else\lslash\fi}

\newcommand\beq{\begin{equation}}
\newcommand\eeq{\end{equation}}
\newcommand\dst{\displaystyle}

\newcommand\ssst{\scriptscriptstyle}

\newcommand\Span{{\rm span}}

\let\footnoteIOP\footnote
\renewcommand\footnote[1]{
	\ifthenelse{\value{footnote}=9}{\setcounter{footnote}{1}}{}
	\footnoteIOP{#1}
}

\begin{document}

\title{Relativistic Positioning Systems: The Emission Coordinates}

\newcommand\SYRTE{SYRTE--CNRS,
	Observatoire de Paris.
	61, Avenue de l'Observatoire.
	F-75014 Paris.}
\newcommand\FFN{Departament de F\'{\i}sica Fonamental,
	Universitat de Barcelona. Diagonal, 647. E-08028 Barcelona.}

\author{Bartolom\'e Coll$^1$ and Jos\'e M Pozo$^{1,2}$}
\address{$^1$ SYRTE, CNRS -- Observatoire de Paris, 61, Avenue de l'Observatoire, F-75014, Paris, France}
\address{$^2$ Departament de F\'{\i}sica Fonamental, Universitat de Barcelona. Av. Diagonal, 647, E-08028 Barcelona}
\eads{\mailto{bartolome.coll@obspm.fr}, \mailto{Jose-Maria.Pozo@obspm.fr}}

\begin{abstract}
This paper introduces some general properties of the gravitational metric and the natural basis of vectors and covectors in 4-dimensional emission coordinates. Emission coordinates are a class of space-time coordinates defined and generated by 4 emitters (satellites) broadcasting their proper time by means of electromagnetic signals. They are a constitutive ingredient of the simplest conceivable relativistic positioning systems. Their study is aimed to develop a theory of these positioning systems, based on the framework and concepts of general relativity, as opposed to introducing `relativistic effects' in a classical framework. In particular, we characterize the causal character of the coordinate vectors, covectors and 2-planes, which are of an unusual type. We obtain the inequality conditions for the contravariant metric to be Lorentzian, and the non-trivial and unexpected identities satisfied by the angles formed by each pair of natural vectors.  We also prove that the metric can be naturally split in such a way that there appear 2 parameters (scalar functions) dependent exclusively on the trajectory of the emitters, hence independent of the time broadcast, and 4 parameters, one for each emitter, scaling linearly with the time broadcast by the corresponding satellite, hence independent of the others.
\end{abstract}


\submitto{\CQG}

\maketitle

\section{Introduction}

A system of coordinates for some domain of a four-dimensional space-time may be given by a set of four scalar fields, provided with the required properties of continuity, differentiability and regularity (non-degeneration) in the considered domain. If the four scalar fields are realized as direct measures of physical fields, we say that they constitute a system of {\em physical} coordinates. The most simple example of physical coordinates is given by four emitters $S_A$, each of them broadcasting a time scale $\sigma^A$ by means of electromagnetic signals. In particular, we have in mind the most natural case, in which the time scale is based on the proper time $\tau^A$ of every emitter.

We suppose here physical situations in which the broadcasting  electromagnetic signals may be correctly described by ideal electromagnetic signals propagating along null geodesics (electromagnetic optical approximation in vacuum). Thus, the set of space-time events determined (or labeled) by each value of the signal is the future-oriented null cone with vertex on the emitter world-line at the moment when the signal is emitted. Consequently, the four (one-parameter families of) coordinate hypersurfaces $\tau^A={}$constant
are null. This implies that the four one-forms $d\tau^A$, defining the natural or coordinate coframe\footnote{As usual, the word {\em coframe} is used as a shortcut for a frame of covectors.} $\{d\tau^A\}$ of the coordinate system $\{\tau^A\}$, are all null. Such a coframe is of a very unusual causal class\footnote{For the classification of all the 199 causal classes of frames of the space-time see \cite{coll-morales}. For some coordinate examples see \cite{Morales}.}

To such a coframe of null covectors corresponds the metrically associated frame of null vectors  $(\vec\ell^A)^\mu=g^{\mu\nu}(d\tau^A)_\nu$ determining the null geodesics along which the signals propagate. Notice that neither  these vectors nor its directions coincide with any of the vectors or the directions of the dual basis: $\vec\ell^A\neq \partial_A$. Indeed, as we will see, the natural or coordinate vectors, $\partial_A$, of these coordinates are not null but all space-like. This class of  coordinates will be called {\em emission coordinates}. This is a sound name since in this construction the coordinates themselves are broadcast by the four emitter world-lines. Accordingly, a frame of four future oriented null covectors will be called an {\em emission frame}.

Let us observe that the appellation {\em null coframe} is also very  appropriate for an emission frame $\{d\tau^A\}$, because describing the causal orientation of {\em all} its real constituents, the covectors $d\tau^A.$ Nevertheless, attention must be paid to the context in which it is used, because this appellation has been also frequently applied to the very different Newman-Penrose frames \cite{N-P}, both in their real and complex versions.

In fact, this usage has induced some misconceptions about the existence of real frames of real null vectors or covectors in a space-time of hyperbolic signature. The point is that for Newman-Penrose frames the real pair of null vectors is chosen  orthogonal to the remaining vectors of the frame, preventing these last ones to be also real and null. And it is this same orthogonality choice which imposes that the dual frame contain similarly a pair of real null vectors.
The absence of this orthogonality constraint in emission frames, because of its absence of physical role, allows these frames, on one hand, to be constituted by four real null covectors, and forbids their dual frames, on the other hand, to contain real null vectors. In fact, as commented above, all the vectors of the dual frames are always  space-like.

Real null frames seem to have been first considered by Zeeman \cite{Zeeman} as a device for a technical proof. Derrick \cite{Derrick81} discovered a class of them as particular {\em symmetric frames}, later studied by Coll and Morales \cite{Coll&Morales91}, who also proved that real null frames constitute a causal class among the 199 possible ones \cite{coll-morales}. Coll \cite{Yoluz} seems to have been the first to construct physical coordinate
systems by means of light beams. The real null frames associated to light beams are, in some sense, dual to emission frames: they are the natural vectors which are null in this case, while the covectors are space-like. 
Symmetric real null frames have also been proposed by Finkelstein and Gibbs \cite{Finkelstein}  as a checkerboard lattice for a quantum space-time. 
It is also Coll \cite{Coll01, collbrussels, coll-1} which seems the first to have been proposed the physical construction of relativistic coordinate systems by means of broadcast light signals, whose natural coframe is an emission frame. 
Bahder \cite{bahder} has obtained explicit calculations for the
vicinity of the Earth at first order in the Schwarzschild
space-time, and Rovelli \cite{Rovelli02} has developped a particular case
where emitters define a symmetric frame in Minkowski space-time, as representative of a complete set of gauge invariant observables.
Blagojevi\`c et al. \cite{Blagojevic02} analysed and developped the
symmetric frame considered in Finkelstein and Rovelli papers.
Recently Lachi\`eze-Rey \cite{LachiezeRey} has considered applications of emission coordinates to cosmology and positioning. Some more specific papers on positioning systems have been published \cite{Coll2004,ColTar,pozo,pozoERE05,moralesERE05,ferrandoERE05,yoERE05,2LPhysRev}, an international school has been devoted to the subject \cite{salamanca}, and some works are in progress \cite{CFM2006b,CollPozo06b,CollPozo06c,CollPozo06d}.

The interest of emission coordinates lies in that they are constitutive ingredients of (relativistic) {\em positioning systems.} What is, in turn, the interest of these systems?

In a fully relativistic theory of {\em location systems,} that is in a theory entirely  based on the framework and concepts of general relativity and concerning the physical construction of coordinate systems, relativistic positioning systems appear as the {\em best} systems that today we are able to conceive (see for example \cite{yoERE05}).

General Relativity may be either just applied as a learned algorithm to sprinkle Newtonian expressions with corrective 'relativistic effects' (post-Newtonian perturbation methods) or considered as providing the best concepts on space-time and gravitation.

But today we know that the physical model of space-time and gravitation offered by General Relativity is better adapted to Nature than the model offered by Newtonian theory. As a consequence, Newtonian analysis of global navigation systems, space physics or Solar system astronomy, for example, need more and more of relativistic corrections.

In this situation, the interest of a complete relativistic approach to physical problems, naturally integrating in their starting quantities all the relativistic corrections, without making them explicit as perturbations with respect to an insufficient (and incorrect!) Newtonian theory, seems evident\footnote{In fact, many relativistic analysis and descriptions of physical  systems, including basic properties of the electromagnetic field, shock waves, hydrodynamics and magnetohydrodynamics, detonation and deflagration waves, are much easier to make in relativity than in Newtonian theory.  Contrarily to an extended opinion, they are the high precision quantitative developments searched for some physical systems which makes relativistic calculations long and complex, meanwhile the starting relativistic descriptions of these same physical systems remain relatively simple. Only in the numerical computation of these high precision results, but not in the conceptual context of defining and analyzing the corresponding physical systems, Newtonian theory could be accepted today as the zeroth order term of a numerical algorithm.}.
Aged of almost a century, it is time to  consider the infancy of relativity, paternally guided by Newtonian theory, as finished. 
It is obvious that it is not taking refuge in out-of-date Newtonian concepts that one will be able, there were relativity is concerned by its own right, to ask vanguard scientific questions. We believe that General Relativity theory is ripe enough, and that the present moment is technically and scientifically interesting enough, to develop General Relativity in  an adult, autonomous, form, without reference to Newtonian theory\footnote{We believe that relativity has not still revealed all its capacities of amazing us. Nevertheless, apart from the structural or mathematical generalizations that it suggest, its heuristic power is at present deflated by the oppressive presence of Newtonian theory in almost all its applications. We hope that an autonomous development of General Relativity for physical applications, starting from its proper basis and excluding any a priori help of Newtonian conceptions, may reveal new specific features of conceptual, scientific and technical interest. First results in this direction may be considered those already obtained elsewhere, concerning the new, paradigmatic, way of using GPS satellites as the principal reference system for the Earth (project SYPOR, see \cite{yoERE05}), or the way to use the signals of four millisecond pulsars as a coordinate system for the Solar system and its neighbors (see \cite{ColTar}).}. 

The first obstruction to such an autonomous development of General Relativity is the almost absolute lack of operational protocols for constructing coordinate systems. 
On one hand this lack is due to the fact that in Newtonian physics such protocols are supposed well known and trivial and that, as above mentioned, relativity is frequently used to find relativistic corrections to the Newtonian picture of the physical system in question. On the other hand, it is also due to a misunderstanding of the meaning of the relativistic covariance principle, frequently interpreted erroneously as stating the lack of physical interest of coordinate systems\footnote{The covariance principle is one of the general principles in Physics that help us to better understand the internal structure of physical laws, allowing to express them in easy and clear form. It is an extension of the other one of dimensional invariance, which  states that the physical laws are independent of the particular units used to obtain them. But, can we infer from these principles that the tasks of definition and construction of coordinate systems or units are not of physical interest? Certainly not. The importance of these principles lies in what they allow {\em  separating} the difficulties inherent to the conception and construction of every experiment and to the control of the context in which it takes place, from the difficulties inherent to the conception and construction of the coordinate systems for its location and of the units for the measure of its specific quantities. It would be a dangerous nonsense for physics to believe that, instead of simply allowing this separation, these principles of covariance and of dimensional invariance scorn the physical construction of coordinate systems and units. Far from that, these principles reinforce the need to {\em first} improve the construction of coordinate systems and units as one efficient way to {\em afterward} improve physical laws.}. 

 This is why the first step for a autonomous relativistic analysis of physical situations is the elaboration of a fully relativistic theory of location systems. Its basic ingredients have been obtained elsewhere (see for example \cite{yoERE05} or \cite{2LPhysRev}). It appear that relativistic positioning systems are the best of the physical realizations of coordinate systems conceived up today, because of their {\em immediate} character \cite{yoERE05,2LPhysRev}, i.e. their capacity of indicating to every event its proper coordinates {\em without delay} (contrarily to what happens with {\em all} the other location systems considered up today). 

We are here concerned with these relativistic positioning systems.
 
Relativistic positioning systems are those particular location systems constituted by four clocks broadcasting their proper times\footnote{Slightly more generally, relativistic positioning systems are constituted by four pointlike sources broadcasting countable electromagnetic signals. It is a such one that can be constructed for the Solar system from the signals emitted by four millisecond pulsars \cite{ColTar}}, the coordinates that they generate being the above described emission coordinates. 

Thus, in order to develop these relativistic positioning systems, it is essential to know the properties of emission coordinates and its relationship with other more usual coordinates, as for instance Minkowskian coordinates in the case of a flat space-time. But it is also central to study the additional measures or data to be supplied to the receivers (in general, by broadcasting them together with the coordinates) to give them the desired information, such as the local metric in these coordinates, or to enable them to calculate the position of the emitters in the coordinates generated by themselves. Many important results in this direction has been already obtained for two-dimensional space-times \cite{2LPhysRev,salamanca}. Evidently, the real case of four-dimensional space-times is much more complex and requires a deeper study.

The paper is devoted to the properties of emission coordinates that are inherited by the natural basis that they induce on the tangent and cotangent spaces at every event, that is to say to the emission frames and coframes associated to these coordinates. We characterize, from the definition of emission coordinates and the properties of four-dimensional space-times, the causal type of the natural vectors, covectors and 2-planes of emission frames. The natural covectors are null and future-directed. This implies a particular form for the contravariant metric: the diagonal components vanish, $g^{AA}=0$, and the extra-diagonal terms are positive, $g^{AB}>0$, and satisfy the inequalities corresponding to the three geometric means  $\sqrt{g^{12}g^{34}}$, $\sqrt{g^{13}g^{24}}$ and $\sqrt{g^{14}g^{23}}$ forming a triangle. Indeed, the determinant of the metric depends only on these three quantities. The natural vectors are all space-like. Moreover, the coordinate 2-surfaces are space-like. This means that each pair of vectors, $\partial_A$, $\partial_B$, of an emission frame defines a space-like plane and forms an angle $\theta_{AB}$ between them. We deduce that the six angles obtained in this way satisfy four identites: 
$\theta_{12}=\theta_{34}$, $\theta_{13}=\theta_{24}$, $\theta_{14}=\theta_{23}$ and $\theta_{12}+\theta_{23}+\theta_{13}=2\pi$, thus they have only 2 degrees of freedom. These results involve the characterization of the covariant and contravariant gravitational metric in emission coordinates and lead us to a natural, physically meaningful, normalization of emission frames. In turn, this results into a interesting splitting of the six degrees of freedom of the metric into two clearly different types of parameters: Four {\em scaling parameters}, each corresponding to each emitter, dependent linearly on the proper time transfer of its corresponding clock and independent of the other clocks. And two {\em scale-invariant parameters}, corresponding to the 2 degrees of freedom of the angles $\theta_{AB}$, depending exclusively on the relative directions of the clocks, determined by the world-line of the emitters, independently of the frequency of the emitted series of signals generated by the clocks. This last result has been presented without proof in \cite{pozo,pozoERE05}. For comparison, we present the analogous splitting for three-dimensional space-times. In this case the three degrees of freedom of the metric would be split into three scaling parameters and no scale-invariant parameter.

The properties of emission frames are obtained by using known geometric results for space-times, that is for four-dimensional metric spaces of Lorentzian signature. Thus, our presentation stresses the geometry of the problem by using, when possible, intuitive (and rigorous) deductions rather than algebra. Nevertheless, some of them are afterwards algebraically checked.

Mathematically, the relations obtained in the present work are algebraic functions of the corresponding emission frames and coframes and thus they are first order differential in the coordinates themselves. This last fact will be used in a next work to analyze the  precise conditions for continuity, differentiability and regularity of emission coordinates.
 
The study undertaken in the present paper are also unavoidable in order to tackle  local and global features of emission coordinates and their relationship with more usual coordinates. These features will depend on the space-time considered and on the world-line of the emitters generating the coordinates. Elsewhere we will present  \cite{CollPozo06b,CollPozo06c} the properties obtained for static and stationary positioning systems in flat Minkowski space-time.

The paper is organized as follows. In section \ref{sec:Description} we define emission coordinates and describe the physical situation in which they appear. In section \ref{sec:NaturalCobasis} we study the natural coframe and the corresponding contravariant metric. Then, in section \ref{sec:NaturalBasis} we study their counterpart: the natural frame and the corresponding covariant metric. The properties obtained lead to the above mentioned splitting of the metric, presented in section \ref{sec:MetricSplitting}.
Finally, in section \ref{sec:Co&ContraMetrics} we present the relationship between the properties obtained for the covariant and the contravariant forms of the metric, and in section VII we conclude with a brief analysis of the results obtained and a sketch of the work in progress.

\section{Description of the emission coordinates
		\label{sec:Description}}

To provide a region of the space-time with a coordinate system, it is sufficient to give four scalar fields, say $x^\mu$. Then, their equipotential hypersurfaces $x^\mu ={}$ constant, considered as  (one-parameter family of) coordinate hypersurfaces, define  unambiguously the coordinate system. These fields must satisfy some sufficient conditions of smoothness and regularity: each of them need to be continuous and differentiable until some degree,  and the set of them must be non-degenerate, i.e. they must be functionally independent: $dx^1\w \cdots\w dx^4\neq0$.

As physical fields, the four scalar fields have to be easy to be produced, regular in their propagation, sufficiently weak to be considered as {\em passive}\footnote{Here a scalar field is said passive if its interaction with the events to be located may be considered as negligible. See \cite{yoERE05} or \cite{Coll01} for related properties of a physical coordinate system.} and sufficiently strong to be measured by a local receiver in the space-time domain in question. For many applications in space physics and Earth sciences, the best candidates for these scalar fields
are electromagnetic signals, and the simplest way to parameterize them is to submit them to broadcast the (proper) time of a clock. Thus, a paradigmatic representative class of positioning systems are those constituted by four clocks broadcasting their proper time by means of electromagnetic signals.

Physical conditions here are supposed such that every clock with its emission mechanism of electromagnetic signals broadcasting its proper time may be considered as point-like. Accordingly, its history in the space-time is described by a time-like world-line, and its proper time will  be determined by the gravitational field of the space-time in question (the metric) and by the dynamics of its world-line (its acceleration).

Alternatively, we may consider clocks generating non proper time scales, for example because of a non-ideal or insufficiently controlled mechanism, or because of the deliberate design of synchronizing it with some coordinate time. In any case, any non-proper time-scale will be supposed to endow the world-line of the clock with an increasingly monotonic parameter.

Also, the physical conditions here are supposed such that the electromagnetic signals may be correctly described by the electromagnetic optical approximation in vacuum. Under these circumstances, each time signal emitted by the clock will propagate in the space-time following the future null cone with vertex in the world-line of the clock at the instant of emission. This future null cone is determined by the space-time (its conformal structure, indeed) and the emission event. 

An arbitrary clock broadcasting regularly its time by a series of electromagnetic signals under the above physical conditions will be called for short an {\em emitter}.

\begin{figure}[htb]
\centerline{
\includegraphics[width=0.19\textwidth]{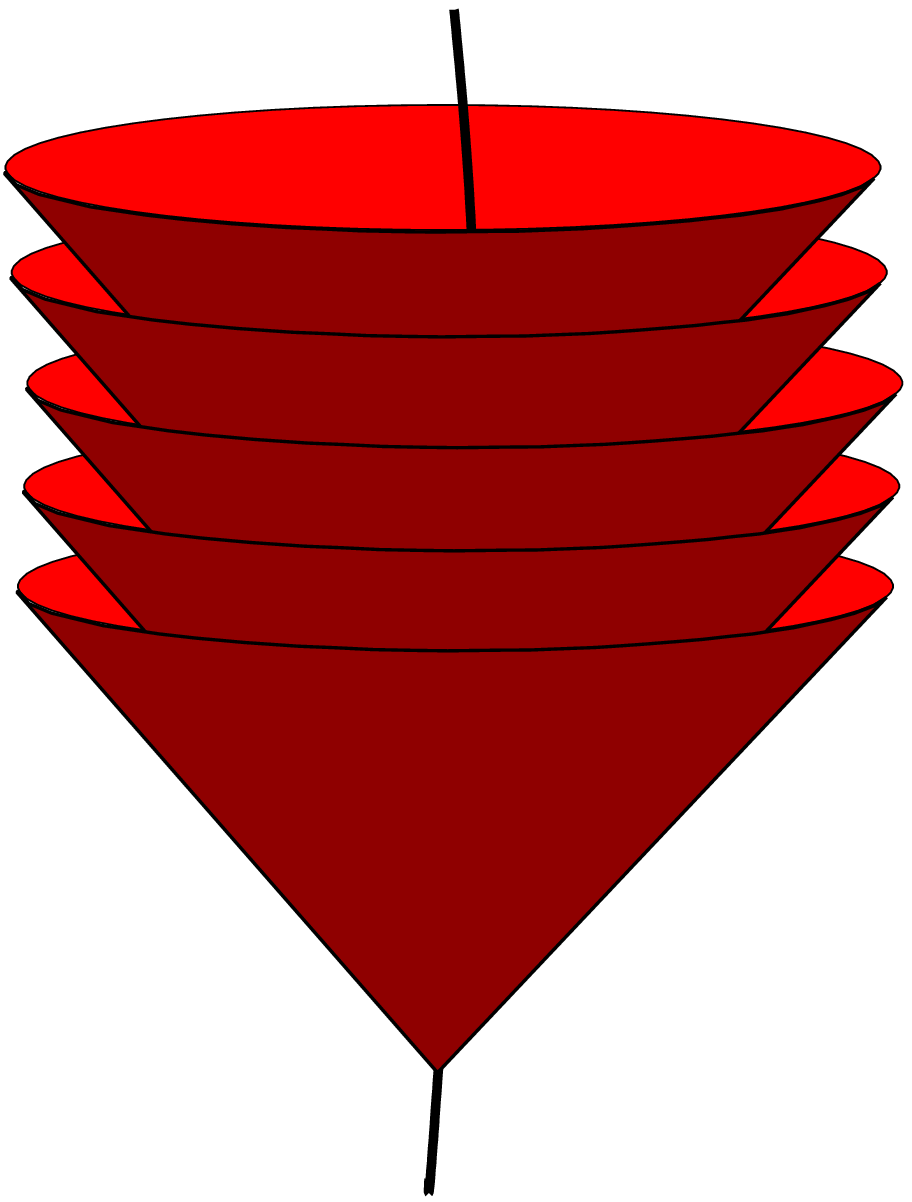}%
\hspace*{0.07\textwidth}
\includegraphics[width=0.42\textwidth]{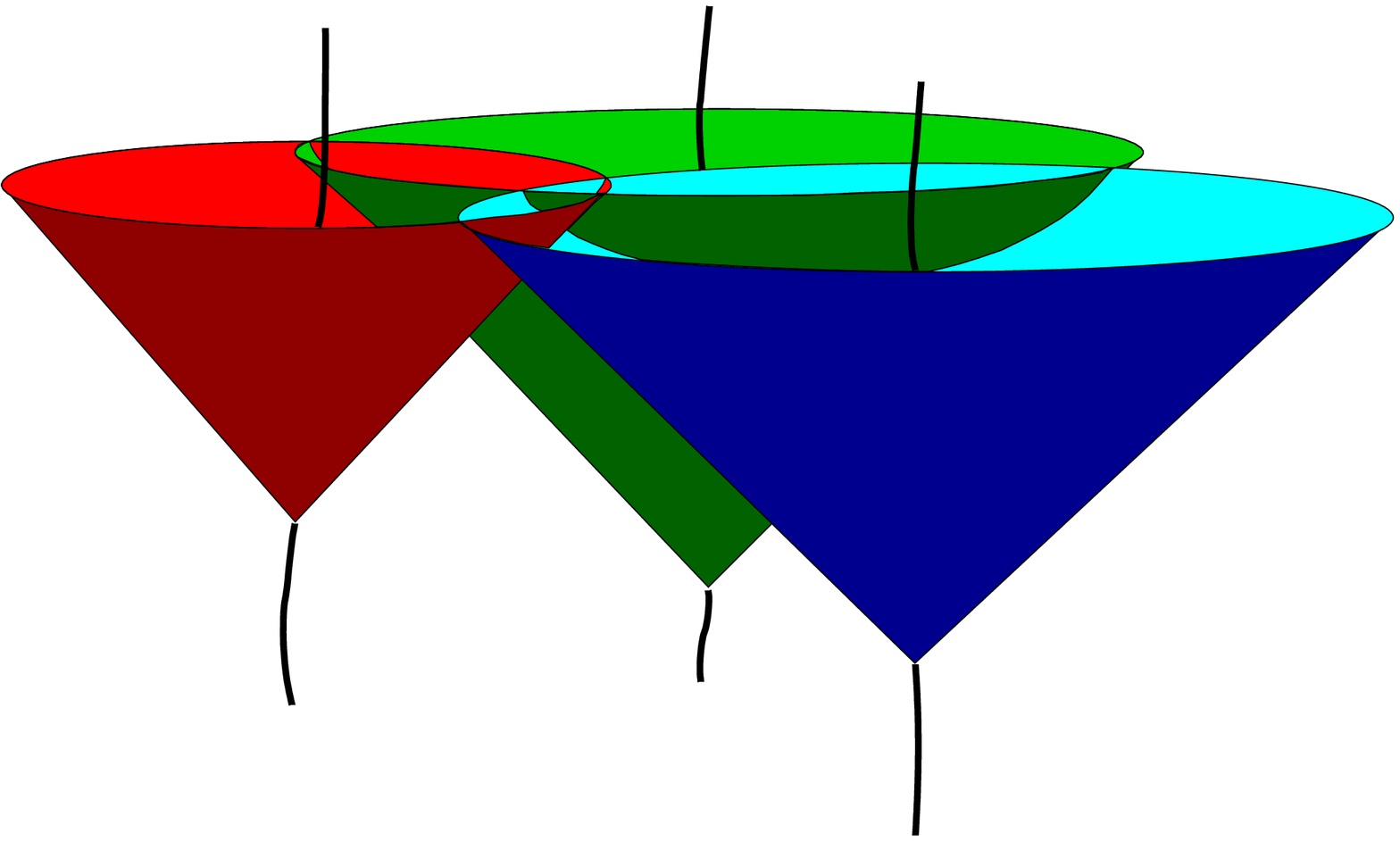}%
\hspace*{0.05\textwidth}
\includegraphics[width=0.22\textwidth]{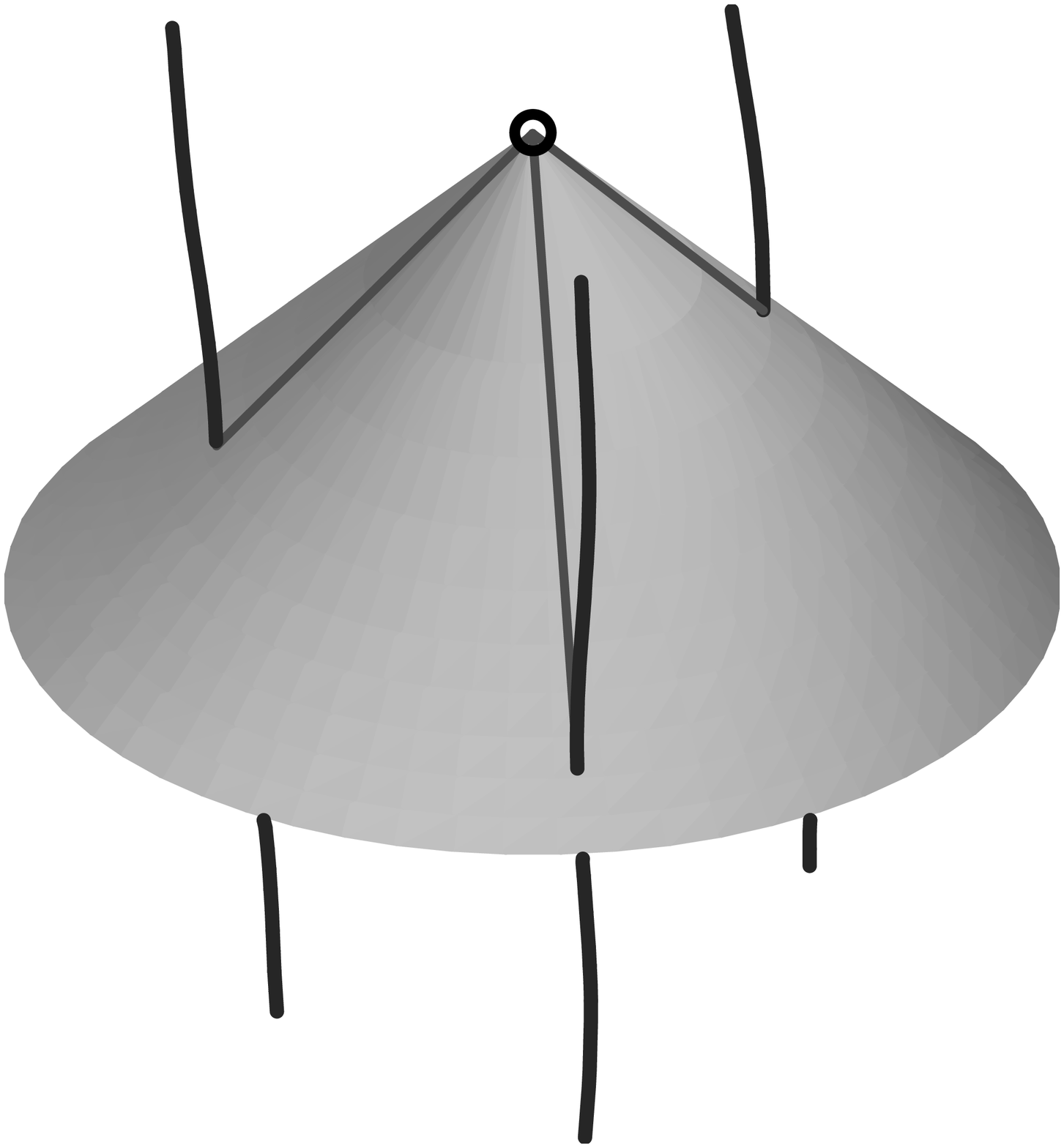}}
\caption{Representation of three emitters in a three-dimensional space-time (time vertical). The lines are the space-time tr ajectories (world-lines) of the emitters. The left figure shows the foliation defined by the space-time surfaces (future light-cones) visited by each value of the signal series of one emitter. The middle figure represents the grid of emission coordinates defined by the series of three  emitters. The right figure gives a complementary view: the intersection of the past light-cone of an event with the trajectories of the emitters gives the emission coordinates of this event.
		\label{fig:NullHypersurfaces}}
\end{figure}

Hence, an emitter defines a family of null cones, which foliates the space time. Each null cone is characterized by the time of its emission and, by interpolation, this defines an scalar field. Then, four different emitters will define four scalar fields which can constitute a coordinate system for some region of the space-time (see figure \ref{fig:NullHypersurfaces}). This class of coordinates, generated by these particular families of hypersurfaces (light cones), are called, as already said, emission coordinates. 

In emission coordinates, the coordinates of an arbitrary event are the four times of emission, carried by the four null cones of the system, that contain the event.

An alternative and complementary view of this construction, which directly induces a simpler way to compute the emission coordinates of an event, is based on the sole past null cone of the event. This past null cone contains every incoming null geodesic, in particular the four geodesics followed by the signals reaching the event from the four emitters. The intersection of this cone with each of the four world-lines determines the emission coordinates of the event (figure \ref{fig:NullHypersurfaces}).

In a subsequent paper we will approach the conditions for the continuity, differentiability and regularity of emission coordinates. For the regularity (non-degeneracy) of the system we need to study first the properties of the emission frames, that is, the natural frames and co-frames in emission coordinates.

\section{The natural covectors and the contravariant metric in emission coordinates
		\label{sec:NaturalCobasis}}

\subsection{Causal character of the natural covectors}

The ideal electromagnetic signals that carry the values $\tau^A$ defining  emission coordinates propagate at the velocity of light. This means that the 
locus of events characterized by one fixed value of one of the coordinates, say $\tau^A={}$const, is the future null cone with the vertex at the source $A$ when it emitted this value. That is, the coordinate hypersurfaces of emission coordinates are always null. Accordingly, 

\begin{property}
The coordinate covectors of emission coordinates, $d\tau^A$, are null.
\end{property} 

The signals sent by each of the emitters constitute an increasing series of values, $\tau^A$ (paradigmatically, its proper time). Thus, if we consider the hypersurfaces given by $\tau^A=t_1$ and $\tau^A=t_2$, with $t_2>t_1$, the later is a null cone emitted after the former. Hence, the hypersuface $\tau^A=t_2$ is in the causal future of $\tau^A=t_1$. This implies that any future-oriented time-like curve, $C(\l)$, will cross the coordinate hypersurfaces $\tau^A={}$const. in increasing order: 
$\frac{d C^\mu}{d\l}\partial_\mu\tau^A>0$. Equivalently, for all time-like and future-directed vector $u$ , the application of the 1-form $d\tau^A$ is positive: $d\tau^A(u)>0$. That is,

\begin{property}
The coordinate covectors of emission coordinates, $d\tau^A$, are future-directed.
\end{property} 

These two properties implies that the metric product of every one of these covectors by itself is null: $d\tau^A\inn d\tau^A=0$, and that, for signature $({+}{-}{-}{-})$, the crossed metric products are positive: 
$d\tau^A\inn d\tau^B>0,\ \forall A,B\neq$. Therefore, the contravariant metric in emission coordinates has the form
\beq
		\label{eq:EmissionContrametric}
	(g^{AB})=\begin{pmatrix}
		0&g^{12}&g^{13}&g^{14} \cr
		g^{12}&0&g^{23}&g^{24} \cr
		g^{13}&g^{23}&0&g^{34} \cr
		g^{14}&g^{24}&g^{34}&0
	\end{pmatrix}
\eeq
where $g^{AB}>0$ for $A\neq B$.

\subsection{Linear independence of the covectors}

Let us denote the natural covectors by $\ell^A\equiv d\tau^A$.
For emission coordinates in a tridimensional space-time, the inequality $g^{AB}>0$ would suffice to guarantee the linear independence of the cobasis $\{\ell^A\}$. In contrast, for the four-dimensional case, this inequality is not sufficient. This difference can be expressed with the following two facts, valid for Lorentzian space-times:
\begin{itemize}

\item Three null vectors are linearly dependent if and only if two 
of the vectors are proportional. That is, three {\em different} null directions 
always span a tridimensional space-time.

\item Four null vectors can be linearly dependent although none of them is
proportional to another. That is, four {\em different} null directions do not
necessarily span a four-dimensional space-time.

\end{itemize}
This is easy to see since there are only two null directions in a bidimensional space-time, but there are infinite null directions in a tridimensional space-time. Thus, we can always choose four null directions living in any tridimensional time-like subspace of the four-dimensional space-time.

The linear dependence of the four covectors $\{\ell^A\}$ is reflected
in the vanishing of the determinant of the metric (\ref{eq:EmissionContrametric}): $|g^{AB}|=0$. Hence,
the condition $|g^{AB}|\neq 0$ implies that the covectors are linearly 
independent.

\subsection{Lorentzian signature of the metric}

The condition $|g^{AB}|\neq 0$ is required in order to have a regular or non-degenerate  metric. However this condition does not ensure that the metric is of
Lorentzian signature, which is required by the space-time properties. 
This requirement introduce another qualitative difference between the 
tridimensional and the four-dimensional case.

\begin{itemize}

\item The existence of null covectors forbids the Euclidean signatures in both
dimensions: $({+}{+}{+})$, $({-}{-}{-})$, $({+}{+}{+}{+})$ and $({-}{-}{-}{-})$. 

\item The Lorentzian signatures $({-}{+}{+})$ and $({-}{+}{+}{+})$, in three and four dimensions respectively, are forbidden by the condition $\ell^A\inn\ell^B>0$, since, in this case, it implies that $\ell^A$ has opposite time orientation to 
$\ell^B$, being impossible to have three covectors, $\ell^1,\ell^2,\ell^3$, with mutually opposite orientation.
\end{itemize}
Therefore:
\begin{itemize}
\item Three null covectors, $\{\ell^A\}$, with positive scalar product for each
pair, $\ell^A\inn\ell^B>0$ ($A\neq B$), span a tridimensional space of
Lorentzian signature $({+}{-}{-})$.

\item Four {\em linearly independent }null covectors, $\{\ell^A\}$, with positive scalar product for each pair, $\ell^A\inn\ell^B>0$ ($A\neq B$), span a four-dimensional space of either Lorentzian signature $({+}{-}{-}{-})$ or null signature $({+}{+}{-}{-})$.

\end{itemize}
This means that, in three dimensions, the Lorentzian signature $({+}{-}{-})$ of $g^{AB}$
is automatically satisfied. This can be checked by observing that the 
determinant of the metric is positive:
\[
	\det(g^{AB})=\begin{vmatrix}
		0&g^{12}&g^{13} \cr
		g^{12}&0&g^{23} \cr
		g^{13}&g^{23}&0
	\end{vmatrix}=2g^{12}g^{13}g^{23}>0\ .
\]
However, in four dimensions, in order to select only the Lorentzian signature,
$({+}{-}{-}{-})$, we need to impose the additional condition $\det(g^{AB})<0$. Indeed, from the above discussion, this condition is necessary and sufficient. 

Let us compute the determinant of the metric. After some manipulation it can be obtained a very interesting factorization of the determinant:
\begin{eqnarray}
	\fl
	|g^{AB}| = \begin{vmatrix}
		0&g^{12}&g^{13}&g^{14} \cr
		g^{12}&0&g^{23}&g^{24} \cr
		g^{13}&g^{23}&0&g^{34} \cr
		g^{14}&g^{24}&g^{34}&0
	\end{vmatrix}
	&=A^4+B^4+C^4-2A^2B^2-2A^2C^2-2B^2C^2 \nonumber
	\\ 
	&=(A+B+C)(A-B-C)(B-A-C)(C-A-B)
\end{eqnarray}
where $A,$ $B,$ $C$ are the geometric means of the three products of complementary components:
\beq
		\label{eq:ABCdef}
	A\equiv \sqrt{g^{23}g^{14}}\ ,\quad 
	B\equiv \sqrt{g^{13}g^{24}}\ ,\quad 
	C\equiv \sqrt{g^{12}g^{34}}\ .
\eeq
Thus, these three positive parameters, $A,B,C>0$, are the appropriate ones to 
express the conditions for the metric to have non-degenerate Lorentzian 
signature:
\begin{eqnarray}
		\label{eq:TriangularCond}
	\fl
	\det(g^{AB})<0 
	&\iff (B+C-A)(A+C-B)(A+B-C)>0 \nonumber \\
	&\iff  A<B+C\ ,\quad B<A+C\quad\mbox{and}\quad C<A+C\ .
\end{eqnarray}
That is, 
\begin{theorem}
The metric of the form {\em(\ref{eq:EmissionContrametric})} is a non-degenerate Lorentzian metric if and only if the geometric means $A,B,C$ defined by {\em (\ref{eq:ABCdef})} can be the lengths of the sides of a triangle {\em (\ref{eq:TriangularCond})}. 
\end{theorem}

This image is more evident if we realize that the determinant of the metric is minus the Heron polynomial:
\beq
		\label{eq:HeronPolynomial}
	\fl
	-\det(g^{AB})={\cal H}(A,B,C)
	\equiv (A+B+C)(B+C-A)(A+C-B)(A+B-C).
\eeq
Let us recall that the Heron polynomial gives 
the area of a triangle in terms of the length of its sides $A,B,C$:
\[
	\mathop{\mbox{Area}}\,
	\raisebox{1pt}{$\dst {}^{\ssst A\!}
	{\ssst\underset{\ssst C}{\dst\triangle}}
	{}^{\ssst\! B}$}
	=\frac14\sqrt{{\cal H}(A,B,C)} \,.
\]

\section{The natural vectors and the covariant metric in emission coordinates
		\label{sec:NaturalBasis}}

Once we have the contravariant metric, $g^{AB}$, the covariant metric, $g_{AB}$, is obtained simply by inverting the matrix of their components: 
$(g_{AB})=(g^{AB})^{-1}$. However this method does not make evident the geometric properties that the covariant metric must satisfy, which, as for the contravariant metric, are consequence of the Lorentzian signature of the space-time and of the null and future-directed nature of the covectors $d\tau^A$. For this reason we are going to derive the properties of the covariant metric in emission coordinates by means of some simple geometric deductions from well-known properties of Lorentzian spaces. This procedure lead us to obtain a natural splitting of the covariant metric. In the next section we will obtain the relationship between the ingredients of both metrics by the inversion of one of them.

\subsection{Causal character of the natural vectors}

We have seen that the natural covectors, $d\tau^A$, of emission coordinates are null.  However, the natural vectors, $\partial_{\tau^A}$, of the dual frame do not have the same causal character (except for bidimensional space-times). The causal character of the covector $d\tau^A$ is given by the nature of the coordinate hypersurface $\tau^A={}$const. In contrast, the causal character of the vector $\partial_{\tau^A}$ is given by the nature of the coordinate line $\tau^B={}$const. $\forall B\neq A$.
The other object that characterizes from the causal point of view the coordinates is the causal nature of the coordinate 2-surfaces (see ref. \cite{coll-morales}), which can be represented by the bivectors 
$\partial_{\tau^A}\w\partial_{\tau^B}.$
	
The coordinate lines of emission coordinates are the intersection of three null hypersurfaces (light cones). In a null hypersurface, all the directions are space-like except one which is null. Hence, the coordinate lines must be either space-like or null. In addition, at any point of the null hypersurface, the tangent hyperplane is determined by the null direction. Thus, any two null hypersurfaces containing the same null direction at some common point are tangent at this point. Therefore, the coordinate line can be null only where the three hypersurfaces are tangent, i.e. where the coordinate system is degenerate. Consequently:

\begin{property}
The coordinate vectors $\partial_{\tau^A}$ of emission coordinates are space-like.
\end{property}

An alternative way for arriving to this conclusion is the following. Let us consider the null vectors $\vec\ell^A$ metrically associated to the null covectors $\ell^A$, denoted with indices as $(\vec\ell^A)^\mu=g^{\mu\nu}(\ell^A)_\nu$. 
Each coordinate vector, $s_A\equiv\partial_{\tau^A}$, is perpendicular to the three-space spanned by the complementary vectors in the reciprocal basis:
\[
	s_A\perp {\cal S}_A\equiv\Span(\vec\ell^B,\vec\ell^C,\vec\ell^D)\ ,
	\quad\mbox{with }A,B,C,D\neq\footnotemark\ .
\]
\footnotetext{After a set of objects, the symbol $\neq$ means here that the objects are pair-wise different.}
The three-space ${\cal S}_A$ is time-like since, by definition, it contains more than one null direction. Hence, the orthogonal vector $s_A$ is space-like.

This type of argument can also be used to obtain the causal character of the coordinate bivectors, $s_A\w s_B$, that is, the causal character of the plane spanned by each pair of vectors $s_A,s_B$. This plane is completely orthogonal to the plane spanned by the dual vectors $\vec\ell^C,\vec\ell^D$, with $A,B,C,D\neq$. This latter is time-like, since it contains two null directions. Therefore:

\begin{property}
		\label{theo:SpacelikePlanes}
The planes spanned by each pair of coordinate vectors of emission coordinates, $s_A\w s_B$, are space-like.
\end{property}

The three-space spanned by each triad of vectors, 
\[
	{\cal S}^D\equiv\Span(s_A,s_B,s_C)\ ,\quad\mbox{with }A,B,C,D\neq\ ,
\]
is orthogonal to the complementary dual vector $\vec\ell^D$, which is null. Hence, although the three vectors are space-like, they span a null space.

\begin{property}
The three-spaces spanned by each triad of coordinate vectors of emission coordinates, $s_A\w s_B\w s_C$, are null. Their unique null direction is given by the vector $\vec\ell^D$ ($A,B,C,D\neq$).
\end{property}

These last three properties may also be obtained by means of more general but harder methods developped in \cite{coll-morales}.

\subsection{The angles between the natural vectors}

The space ${\cal S}^D$ is degenerated, so that any metric quantity will not depend on the null direction $\vec\ell^D$. We can then consider its quotient space by the null direction, ${\cal S}^D/\vec\ell^D$, which is a bidimensional space with an (anti)-Euclidean metric induced. Let us denote the equivalence class of the vector $v\in{\cal S}^D$ by $\langle v \rangle\in{\cal S}^D/\vec\ell^D$. The {\em non-oriented} angle formed by two vectors is the positive angle given by the metric:
\[
	s_A\inn s_B=-|s_A|\,|s_B|\cos\theta_{AB}\ ,\quad\mbox{with }
	0\leq\theta_{AB}\leq\pi \,.
\]
Note that this last condition implies $\theta_{AB} = \theta_{BA}.$The same angle is also well defined for the equivalence classes, so that $\theta_{AB}$ is also the angle between 
$\langle s_A\rangle,\langle s_B\rangle\in{\cal S}^D/\vec\ell^D$. But 
the three vectors $\langle s_A\rangle,\langle s_B\rangle,\langle s_C\rangle$ are in a bidimensional Euclidian space. This provides the following

\begin{lemma}		\label{theo:NonOrientedAngles}
The non-oriented angles, $\theta_{AB},\theta_{BC},\theta_{AC}$, formed by three space-like vectors $s_A,s_B,s_C$ in a null tridimensional space, 
${\cal S}^D$, satisfy either
\[
	\theta_{AB}+\theta_{BC}+\theta_{CA}=2\pi
	\qquad\mbox{or}\qquad
	\theta_{AB}+\theta_{BC}=\theta_{CA} 
\]
for some angle, $\theta_{CA}$, of them.
\end{lemma}

\begin{figure}[htb]
\centerline{\includegraphics[width=0.67\textwidth]{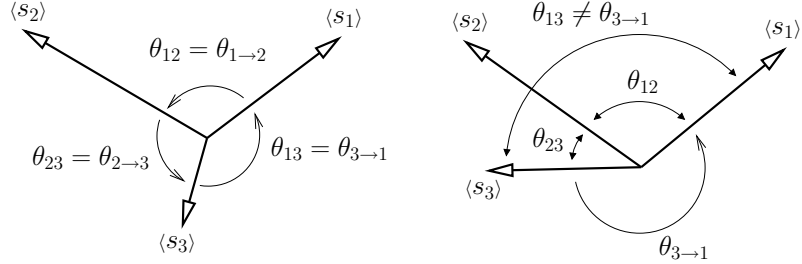}}
\caption{	\label{fig:3AnglesIn2d}
Angles between three vectors in an Euclidian bidimensional space. If the angles are considered oriented then the sum of the three angles is a complete revolution ($\theta_{1\to2}+\theta_{2\to3}+\theta_{3\to1}=2\pi$). But if we consider non-oriented angles ($0\leq\theta_{AB}\leq\pi$), this is only true if the origin is inside the triangle formed by the end points of the vectors. In any other case there will be an oriented angle bigger than $\pi$, say $\theta_{3\to1}>\pi$, so that when considering non-oriented angles the identity satisfied will be $\theta_{13}=\theta_{12}+\theta_{23}$.}
\end{figure}

The different possibilities depend on the disposition of the vectors in the plane (see figure \ref{fig:3AnglesIn2d}).

The null vector $\vec\ell^D$ belongs to the space ${\cal S}^D$ so that it can be spanned by the three vectors:
\beq
		\label{eq:NullCombination}
	\vec\ell^D=\alpha s_A+\beta s_B+\gamma s_C\quad
	\mbox{for some }\alpha,\beta,\gamma\in\R \,.
\eeq
For an emission tetrad the four null vectors $\vec\ell^D$ are future-oriented. This implies that the three coefficients are positive, for instance:
\[
	\ell^A(\vec\ell^D)=\alpha>0\,.
\]
Taking now the equivalence classes of equation (\ref{eq:NullCombination}), we obtain
\[
	0=\alpha \langle s_A\rangle + \beta \langle s_B\rangle 
	+ \gamma \langle s_C\rangle\ ,\quad\mbox{with }\alpha,\beta,\gamma> 0\,,
\]
which implies that the origin is inside the triangle with vertices at the end of the three vectors. Thus, the future orientation select one of the possibilities of lemma \ref{theo:NonOrientedAngles}.

\begin{lemma}	
Let $\{\ell^A\}$ be an emission tetrad and $\{s_A\}$ its dual basis. The non-oriented angles, $\theta_{AB},\theta_{BC},\theta_{AC}$, formed by any triad of
vectors $s_A,s_B,s_C$ of the dual basis, satisfy
\[
	\theta_{AB}+\theta_{BC}+\theta_{CA}=2\pi \, .
\]
\end{lemma}

Taking this identity for the four possible triads of vectors, we obtain another identity:
\[
	\left.
	\eqalign{
	+(\theta_{AB}+\theta_{BC}+\theta_{AC}&=2\pi) \cr
	+(\theta_{AB}+\theta_{BD}+\theta_{AD}&=2\pi) \cr
	-(\theta_{AC}+\theta_{CD}+\theta_{AD}&=2\pi) \cr
	-(\theta_{BC}+\theta_{CD}+\theta_{BD}&=2\pi)
	}
	\right\} \Rightarrow \theta_{AB}=\theta_{CD} \,,
\]
for any $A,B,C,D\neq$. Thus the angle formed by any pair of vectors, $s_A,s_B$, coincides with the angle formed by the complementary pair, $s_C,s_D$. 

\begin{theorem}		\label{theo:SumOfAngles}
Let $\{\ell^A\}$ be an arbitrary emission tetrad and $\{s_A\}$ its dual basis. The non-oriented angles formed by the space-like vectors $s_A$ satisfy
\[
	\theta_{12}=\theta_{34}\ ,\quad
	\theta_{13}=\theta_{24}\ ,\quad
	\theta_{23}=\theta_{14}\ ,
\]
and
\beq
		\label{eq:SumOfAngles}
	\theta_{12}+\theta_{13}+\theta_{23}=2\pi \ ,\quad
	0<\theta_{12},\theta_{13},\theta_{23}<\pi \ .
\eeq
\end{theorem}

Let us observe that this coincidence of the angles by pairs is not a trivial condition. These results say that for whatever four linearly independent null wave-fronts:
\begin{itemize}

\item The intersection of three wave-fronts is a space-like line. This gives the causal nature of the four dual vectors.

\item The intersection of 2 wave-fronts is a space-like surface containing 2 of these lines. This gives the causal nature of the 6 planes formed by each pair of dual vectors.

\item The angle formed by any 2 of the lines coincide with the angle formed by the other 2 lines.

\item The sum of the three angles formed by 1 line and each of the other three lines is $2\pi$.

\end{itemize}

\subsection{The length of the natural vectors}

Observe that the last result says that there are only 2 degrees of freedom for the six angles. The other four degrees of freedom of the metric are contained in the {\em lengths} of the four vectors $s_A$,
\[
	\mu_A\equiv\sqrt{- s_A\inn s_A}\,,
\]
which are independent parameters.
These four lengths, $\mu_A$, and the angles, $\theta_{AB}$, (6 independent parameters in total) determine the covariant metric:
\[
	\eqalign{
	g_{AA}&=-\mu_A{}^2 \cr
	g_{AB}&=-\mu_A\mu_B\cos\theta_{AB}
	\quad\mbox{for }A\neq B\,.
	}
\]
Thus,
\[
	(g_{AB})=-\begin{pmatrix}
	\mu_1{}^2 & \mu_1\mu_2 Z 
	& \mu_1\mu_3 Y & \mu_1\mu_4 X \\
	\mu_1\mu_2 Z & \mu_2{}^2
	& \mu_2\mu_3 X & \mu_2\mu_4 Y \\
	\mu_1\mu_3 Y & \mu_2\mu_3 X
	& \mu_3{}^2 & \mu_3\mu_4 Z \\
	\mu_1\mu_4 X & \mu_2\mu_4 Y
	& \mu_3\mu_4 Z & \mu_4{}^2
	\end{pmatrix}
\]
where
\beq
		\label{eq:cosines}
	X\equiv\cos\theta_{23}\ ,\quad
	Y\equiv\cos\theta_{13}\ ,\quad
	Z\equiv\cos\theta_{12}\ ,
\eeq
and these three angles must satisfy the ligature (\ref{eq:SumOfAngles}). This can be equivalently characterized by the conditions
\beq
		\label{eq:AnglesCosinusConditions}
	X^2+Y^2+Z^2-2XYZ=1 \qquad\mbox{and}\qquad
	\left\{\,\eqalign{
	YZ&>X\cr XZ&>Y\cr XY&>Z \,
	}\right.
\eeq
which can be checked (with some effort) to imply the existence of the angles: 
$X^2,Y^2,Z^2<1$, and their ligature (\ref{eq:SumOfAngles}).

\section{The normalized emission frame and the splitting of the metric
		\label{sec:MetricSplitting}}

\subsection{The normalized emission frame}

In the preceding section we have seen that all the four vectors of the dual basis of an emission frame are space-like. And we have also seen that the four lengths of this vectors are independent metric parameters. This fact leads us to define the {\em normalized vectors}:
\beq
		\label{eq:schapsmu}
	\hat s_A\equiv s_A/\mu_A \,.
\eeq
This normalized vectors constitute a normalized basis with the frame-metric
\[
	\hat s_A\inn\hat s_B \equiv \hat g_{AB}\ ,\quad\mbox{where}\quad
	(\hat g_{AB})=
	- \begin{pmatrix}
	1 & Z & Y & X \\
	Z & 1 & X & Y \\
	Y & X & 1 & Z \\
	X & Y & Z & 1
	\end{pmatrix}
\]
which, according to (\ref{eq:AnglesCosinusConditions}), has only two independent parameters.

In terms of the angles, the determinant of the normalized metric gives the result\footnote{
	The determinant of the original (non normalized) metric is
	\[
		\det(g_{AB})=-(2\mu_1\mu_2\mu_3\mu_4
		\sin\theta_{12}\sin\theta_{13}\sin\theta_{23})^2\,,
	\]
	which gives a completely factorized expression.
	}
\[
	\det(\hat g_{AB})=-(2\sin\theta_{12}\sin\theta_{13}\sin\theta_{23})^2\,.
\]

The normalization (\ref{eq:schapsmu}) can be written by means of the 
{\em normalization matriz} $(M_A{}^{\hat B}),$
\[
	(M_A{}^{\hat B})=
	\begin{pmatrix}
	\mu_1 & 0 & 0 & 0 \\
	0 & \mu_2 & 0 & 0 \\
	0 & 0 & \mu_3 & 0 \\
	0 & 0 & 0 & \mu_4
	\end{pmatrix}
\]
and its inverse, $(N_{\hat A}{}^B)=(M_A{}^{\hat B})^{-1}$, expressing the basis transformation as,
\[
	\hat s_{\hat A}=N_{\hat A}{}^B\,s_B\ ,\quad
	s_A=M_A{}^{\hat B}\, \hat s_{\hat B} \,.
\]

The dual of this normalized basis, $\{\hat s_A\}$, gives a kind of {\em normalized} cobasis, $\{\hat \ell^A\}$, where
\[
	\hat \ell^{\hat A}\equiv M_B{}^{\hat A}\, \ell^B \,.
\]
The covectors $\ell^A$ are null, thus it has no meaning to normalize them individually. However, this construction gives a criterion to normalize a set of four null covectors forming an emission tetrad. The frame-metric of this normalized cobasis, $\hat\ell^A\inn\hat\ell^B\equiv\hat g^{AB}$, is
\beq
		\label{eq:gchapconlatinas}
	(\hat g^{AB})=(\hat g_{AB})^{-1}=\frac1{2 a b c}
	\begin{pmatrix}
	0 & c & b & a \\
	c & 0 & a & b \\
	b & a & 0 & c \\
	a & b & c & 0
	\end{pmatrix}
\eeq
where
\[
	a=\sin \theta_{23}\ ,\quad
	b=\sin \theta_{13}\ ,\quad
	c=\sin \theta_{12}\ .
\]
Observe that while in the covariant metric there appears the cosine of the angles (\ref{eq:cosines}), in the contravariant metric there appear the sines.

The conditions for $a,b,c$ can be proved to be equivalent to the equation
\[
	a^4+b^4+c^4-2a^2b^2-2a^2c^2-2b^2c^2+4a^2b^2c^2=0
\]
and the inequalities $a,b,c>0$.
By the appropriate change we can denote
\beq
		\label{eq:gchapconMayusculas}
	(\hat g^{AB})=
	\begin{pmatrix}
	0 & \hat C & \hat B & \hat A \\
	\hat C & 0 & \hat A & \hat B \\
	\hat B & \hat A & 0 & \hat C \\
	\hat A & \hat B & \hat C & 0
	\end{pmatrix}
\eeq
Then, the condition for $\hat g^{AB}$ to be the frame-metric of a normalized emission frame is\footnote{
	This equation can be alternatively computed from the determinant:
	\[
		\det(\hat g^{AB})=\hat A^4+\hat B^4+\hat C^4
		-2\hat A^2\hat B^2-2\hat A^2\hat C^2-2\hat B^2\hat C^2 
		=\frac{-1}{(2abc)^2}=-2\hat A\hat B\hat C\,.
	\]
	}
\beq
		\label{eq:ContravariantConditions}
	\hat A^4+\hat B^4+\hat C^4
	-2\hat A^2\hat B^2-2\hat A^2\hat C^2-2\hat B^2\hat C^2
	+2\hat A\hat B\hat C=0
\eeq
with $\hat A,\hat B,\hat C>0$.

\subsection{The splitting of the metric}

We have arrived to the normalized emission cobasis $\{\hat\ell^A\}$ by the normalization of the vectors in the dual basis, $\{\hat s_A\}$. This normalization is clearly related to the following splitting of the covariant and contravariant metric:
\[
	g_{AB}=M_A{}^{\hat C} \, \hat g_{\hat C \hat D} \, M_B{}^{\hat D}\,.
	\quad\mbox{and}\quad
	g^{AB}=N_{\hat C}{}^A \, \hat g^{\hat C \hat D} \, N_{\hat D}{}^B
\]
Or, written in matrix notation:
\[
	(g_{AB})=
	-\begin{pmatrix}
	\mu_1 & 0 & 0 & 0 \\
	0 & \mu_2 & 0 & 0 \\
	0 & 0 & \mu_3 & 0 \\
	0 & 0 & 0 & \mu_4
	\end{pmatrix}
	\begin{pmatrix}
	1 & Z & Y & X \\
	Z & 1 & X & Y \\
	Y & X & 1 & Z \\
	X & Y & Z & 1
	\end{pmatrix}
	\begin{pmatrix}
	\mu_1 & 0 & 0 & 0 \\
	0 & \mu_2 & 0 & 0 \\
	0 & 0 & \mu_3 & 0 \\
	0 & 0 & 0 & \mu_4
	\end{pmatrix}^{\sf T}
\]
and
\begin{equation}
		\label{eq:ContravariantSplitting}
	(g^{AB})=
	\begin{pmatrix}
	\mu^1 & 0 & 0 & 0 \\
	0 & \mu^2 & 0 & 0 \\
	0 & 0 & \mu^3 & 0 \\
	0 & 0 & 0 & \mu^4
	\end{pmatrix}
	\begin{pmatrix}
	0 & \hat C & \hat B & \hat A \\
	\hat C & 0 & \hat A & \hat B \\
	\hat B & \hat A & 0 & \hat C \\
	\hat A & \hat B & \hat C & 0
	\end{pmatrix}
	\begin{pmatrix}
	\mu^1 & 0 & 0 & 0 \\
	0 & \mu^2 & 0 & 0 \\
	0 & 0 & \mu^3 & 0 \\
	0 & 0 & 0 & \mu^4
	\end{pmatrix}^{\sf T} \nonumber
\end{equation}
where $\mu^A\equiv 1/\mu_A$.

The interesting point of this splitting is that the metric appears expressed in terms of two clearly different types of parameters: four parameters of `scale' ($\mu_1,\mu_2,\mu_3,\mu_4$) and 2 parameters of `shape' ($X,Y,Z$ with the ligature (\ref{eq:AnglesCosinusConditions}) or $\hat A,\hat B,\hat C$ with the ligature (\ref{eq:ContravariantConditions})). They have the following properties:

\begin{itemize}

\item The shape parameters are invariant respect to the rescaling of the covectors $\ell^A=d\tau^A$, that is, invariant respect to the rescaling of the coordinates:
\[
	\tau^A\mapsto \tau'^A=f^A(\tau^A)\, .
\]
Therefore, $\hat A,\hat B,\hat C$ are independent of the particular time scales broadcasted by the satellites; they only depend on the trajectories (world-lines) of the satellites. 

Furthermore, they are invariant respect to conformal transformations of the metric, depending only on the causal structure of the space-time.

\item Each of the scale parameters $\mu^A$ is invariant respect to the rescaling of the rest of covectors $d\tau^B$, $B\neq A$. Thus, it is invariant respect to the particular time scales broadcasted by the rest of satellites, depending only on the clock of the corresponding satellite. Moreover, it is proportional to the rescaling of the corresponding covector $d\tau^A$. This means that, for the change of coordinates representing the deviation of one of the clocks,
\[
	\tau^A\mapsto f(\tau^A)\quad\mbox{and}\quad
	\tau^B\mapsto \tau^B\quad\forall B\neq A\,,
\]
we have
\[
	\mu^A\mapsto \dot{f}\,\mu^A \quad\mbox{and}\quad\mu^B\mapsto \mu^B
	\quad\forall B\neq A\,,
\]
where $\dot{f}$ is the derivative of $f$.

\end{itemize}

Nevertheless, let us note here that, with respect to the change in the trajectory of the emitters, the modification of just one of them will change all the parameters in a non-trivial way.

\subsection{The splitting of the metric in three dimensions}

We could have introduced the same kind of splitting presented here for the four-dimensional emission metrics also for the tridimensional case. The deduction is completely analogous, with even some simplifications. For instance, there are only three angles $\theta_{12},\theta_{13},\theta_{23}$, and they all coincide: $\theta_{AB}=\pi$. The resulting splitting is thus much more simple:
\beq
		\label{eq:3dMetricSplit}
	\eqalign{
	(g_{AB})&=
	\begin{pmatrix}
	\mu_1 & 0 & 0 \\
	0 & \mu_2 & 0 \\
	0 & 0 & \mu_3
	\end{pmatrix}
	\begin{pmatrix}
	-1 & 1 & 1 \\
	1 & -1 & 1 \\
	1 & 1 & -1
	\end{pmatrix}
	\begin{pmatrix}
	\mu_1 & 0 & 0 \\
	0 & \mu_2 & 0 \\
	0 & 0 & \mu_3
	\end{pmatrix}^{\sf T} \cr
	(g^{AB})&= \frac12
	\begin{pmatrix}
	\mu^1 & 0 & 0 \\
	0 & \mu^2 & 0 \\
	0 & 0 & \mu^3
	\end{pmatrix}
	\begin{pmatrix}
	0 & 1 & 1 \\
	1 & 0 & 1 \\
	1 & 1 & 0
	\end{pmatrix}
	\begin{pmatrix}
	\mu^1 & 0 & 0 \\
	0 & \mu^2 & 0 \\
	0 & 0 & \mu^3
	\end{pmatrix}^{\sf T}
	}
\eeq
Observe that there are three scale parameters, given by 
\[
	\mu^A=\sqrt{\frac{2g^{AB}g^{AC}}{g^{BC}}}\ ,\quad\mbox{with }A,B,C\neq\ ,
\]
and no shape parameter. Accordingly, the normalized metric, $\hat g_{AB}$ or $\hat g^{AB}$,  has no degrees of freedom.

\section{The covariant metric from the contravariant metric
		\label{sec:Co&ContraMetrics}}

At this point one is lead to relate the parameters for the metric obtained in the preceding section with the components of the contravariant metric given as the starting point in formula (\ref{eq:EmissionContrametric}) of section \ref{sec:NaturalCobasis}. The best method to obtain this relationship is computing the inverse of the contravariant metric.

\subsection{The inverse of the cotravariant metric}

The covariant metric, $g_{AB}$, is obtained from the contravariant metric, 
$g^{AB}$, by inverting the corresponding matrix. Recalling that the determinant of the contravariant metric is given by the Heron polynomial 
(\ref{eq:HeronPolynomial}),
$\det(g^{AB})=-\cal H(A,B,C)$, of the parameters $A,B,C$ defined by formula (\ref{eq:ABCdef}), the resulting covariant metric is
\[
	(g_{AB})=\frac{1}{{\cal H}(A,B,C)}
	\begin{pmatrix}
		2g^{23}g^{24}g^{34} & \gamma\ g^{34} &
		\beta\ g^{24} & \alpha\ g^{23} \cr
		\gamma\ g^{34} & 2g^{13}g^{14}g^{34} &
		\alpha\ g^{14} & \beta\ g^{13} \cr
		\beta\ g^{24} & \alpha\ g^{14} &
		2g^{12}g^{14}g^{24} & \gamma\ g^{12} \cr
		\alpha\ g^{23} & \beta\ g^{13} &
		\gamma\ g^{12} & 2g^{12}g^{13}g^{23}
	\end{pmatrix}
\]
where $\alpha\equiv A^2-B^2-C^2,\quad\beta\equiv B^2-A^2-C^2,\quad\gamma\equiv C^2-A^2-B^2$. The components can be compactly written as
\beq
		\label{gAA}
	\eqalign{g_{AA}&=\frac{-2}{\cal H}\,g^{BC}g^{BD}g^{CD}
	\cr
	g_{AB}&=\frac1{\cal H}\, g^{CD}(g^{AB}g^{CD}-g^{AC}g^{BD}-g^{AD}g^{BC})
	}
\eeq
for $A,B,C,D\neq$. Observe that this gives the relationship
\beq
		\label{eq:Co-ContrePairsIdentity}
	g_{AB}/g^{CD}=g_{CD}/g^{AB}
\eeq

\subsection{The algebraic confirmation of the geometric properties}

The components of the covariant metric gives the scalar products between
the natural basis of vectors of emission coordinates, 
$g_{AB}=s_A\inn s_B$, 
where $\{s_A\equiv\partial_A\}$ is the dual basis of $\{\ell^A\equiv d\tau^A\}$.
Thus, we can observe in \ref{gAA} that the four vectors are, as expected, space-like:
\[
	s_A\inn s_A=g_{AA}<0
	\quad\forall\, A=1,2,3,4.
\]
Thus, the scale parameters in terms of the components of the contravariant metric are given by:
\beq
		\label{eq:ScaleParameters}
	\mu_A=\sqrt{-g_{AA}}=\sqrt{\frac{2}{\cal H}g^{BC}g^{BD}g^{CD}},
\eeq
with $A,B,C,D\neq$.

We can also obtain the cosine of the angle between each pair of
vectors:
\[
	\cos\theta_{AB}\equiv\frac{-g_{AB}}{\sqrt{-g_{AA}}\sqrt{-g_{BB}}}\ ,
\]
which from equation (\ref{eq:Co-ContrePairsIdentity}) satisfy
\[
	\cos\theta_{AB}=\frac{-{\cal H}\,g_{AB}}{2g^{CD}\sqrt{g^{BC}g^{BD}g^{AC}g^{AD}}}=\cos\theta_{CD} \,.
\]
confirming the equality of the angles between complementary pairs stated in 
theorem \ref{theo:SumOfAngles}.
This gives us the result
\beq
		\label{eq:cosinesFromABC}
	\eqalign{
	X\equiv\cos\theta_{23}&=\cos\theta_{14}=\frac{A^2-B^2-C^2}{2BC}\cr 
	Y\equiv\cos\theta_{13}&=\cos\theta_{24}=\frac{B^2-A^2-C^2}{2AC}\cr 
	Z\equiv\cos\theta_{12}&=\cos\theta_{34}=\frac{C^2-A^2-B^2}{2AB} \ , 
	}
\eeq
which relates the scale invariant parameters $X,Y,Z$ of the preceding section with the components of the contravariant metric.

Observe that (\ref{eq:cosinesFromABC}) is almost the law of cosines for the triangle of sides $A,B,C$, differing only by having the opposite sign. This means that the angles $\theta_{AB}$ are the supplementary of the angles of this triangle:
\[
	\theta_{12}=\pi-\angle{AB}\ ,\quad 
	\theta_{13}=\pi-\angle{AC}\ ,\quad 
	\theta_{23}=\pi-\angle{BC} \ .
\]
Then, from the well-known property for the sum of the angles of a triangle, $\angle{AB}+\angle{AC}+\angle{BC}=\pi$, it follows
\[
	\theta_{12}+\theta_{13}+\theta_{23}=2\pi \ ,
\]
confirming the result of theorem \ref{theo:SumOfAngles}. This can also be confirmed by checking that (\ref{eq:cosinesFromABC}) implies the conditions (\ref{eq:AnglesCosinusConditions}).

\begin{figure}[htb]
	\centerline{\parbox[c]{0.26\textwidth}{\ \\[1.5em]
		\includegraphics[width=0.26\textwidth]{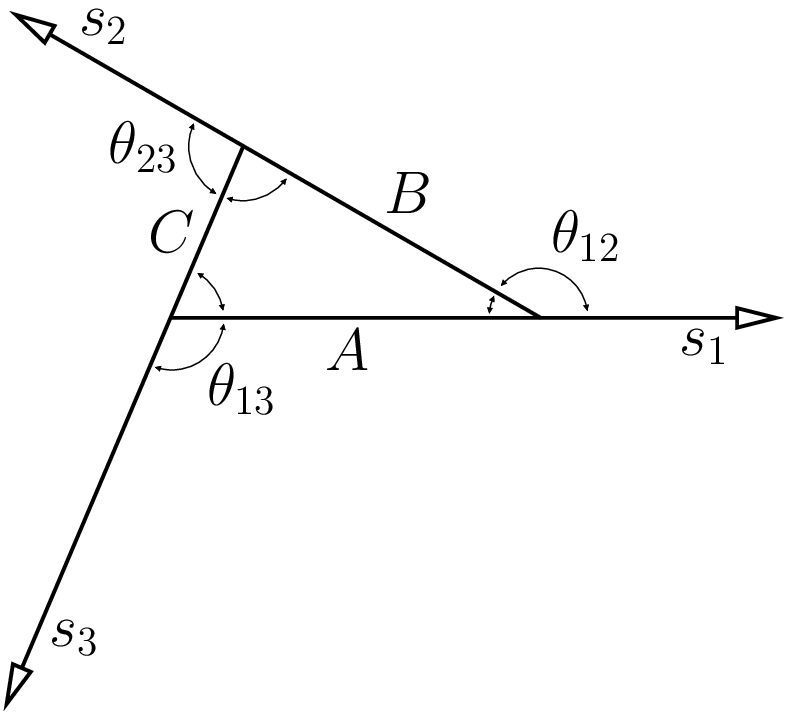}}
	\quad
	\parbox[c]{0.34\textwidth}{
		\includegraphics[width=0.34\textwidth]{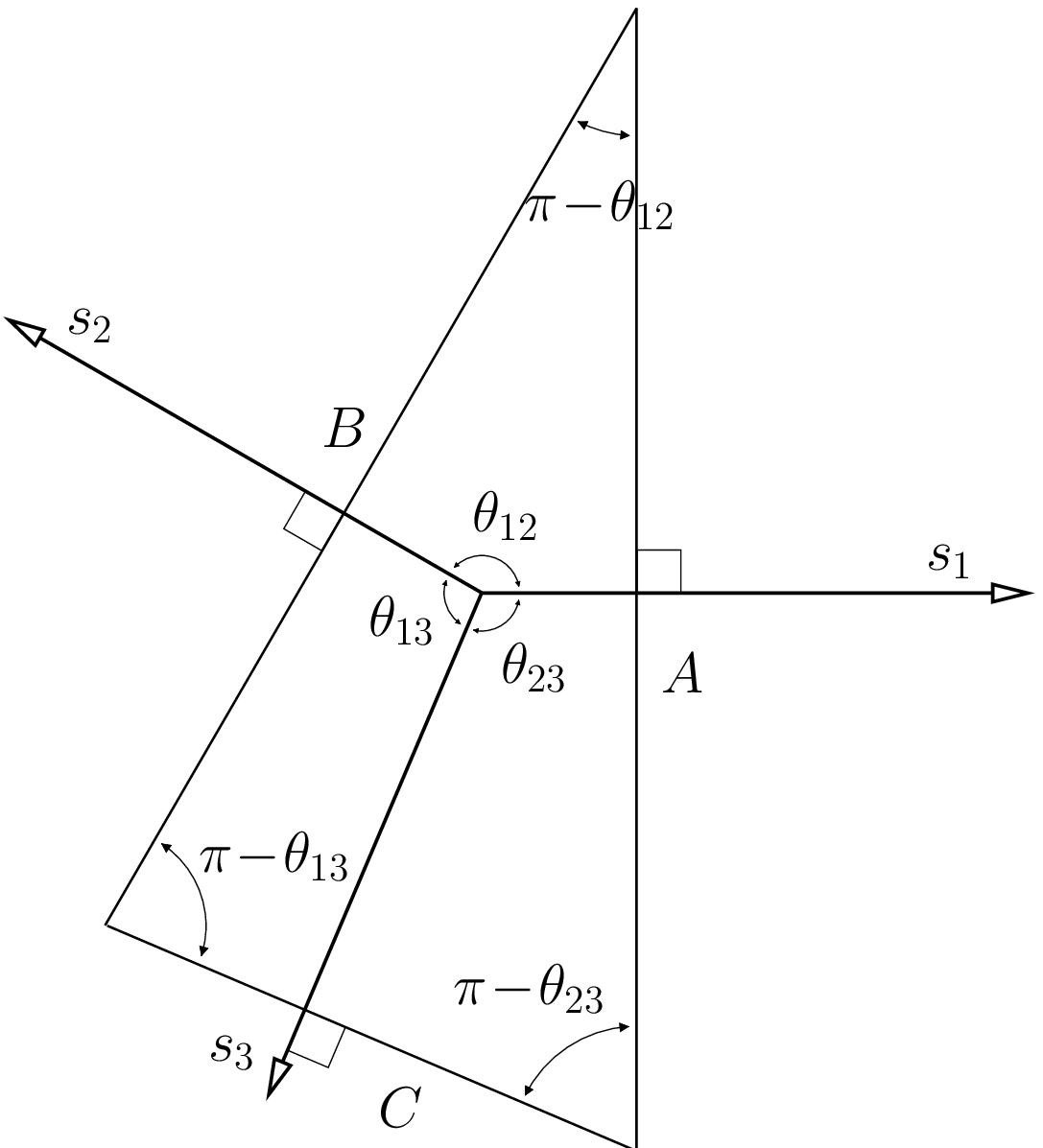}}}
	\caption{Geometrical relationship between the angles formed by the natural
		vectors, $s_A$, and the 	triangle of parameters, $A,B,C$. Two different
		visualizations.
			\label{fig:TriangleOfVectors}}
\end{figure}

\subsection{The scale and shape parameters in terms of the metric components}

In equations (\ref{eq:ScaleParameters}) and (\ref{eq:cosinesFromABC}) we have already found the scale parameters $\mu_A$ and the shape parameters $X,Y,Z$ in terms of the contravariant metric components. But we are also interested in obtaining the expression for the shape parameters $\hat A,\hat B,\hat C$. Let us recall that, from (\ref{eq:gchapconlatinas}) and (\ref{eq:gchapconMayusculas}), they are given in terms of the sines of the angles by
\[
	\hat A=\frac1{2bc} \ ,\quad 
	\hat B=\frac1{2ac} \ ,\quad 
	\hat C=\frac1{2ab} \ ,
\]
where $a=\sin\theta_{23},\ b=\sin\theta_{13},\ c=\sin\theta_{12}$.
Thus by using the identity $\sin\theta_{AB}=\sqrt{1-(\cos\theta_{AB})^2}$ in
(\ref{eq:cosinesFromABC}) we get
\[
	a=\frac{\sqrt{4B^2C^2-(A^2-B^2-C^2)^2}}{2BC}=\frac{\sqrt{\cal H}}{2BC} \ ,\quad
	b=\frac{\sqrt{\cal H}}{2AC} \ ,\quad
	c=\frac{\sqrt{\cal H}}{2AB} \, .
\]
Therefore, we obtain the {\em normalization factor} for the parameters:
\beq
		\label{eq:ShapeParameters}
	\hat A=\frac{2ABC}{\cal H} A\ ,\quad
	\hat B=\frac{2ABC}{\cal H} B\ ,\quad 
	\hat C=\frac{2ABC}{\cal H} C\ .
\eeq

From this expression and equation (\ref{eq:ScaleParameters}) we can check for instance that
$
	g^{23}=\mu^2\mu^3{\hat A}
$,
which confirms the splitting of the contravariant metric given by (\ref{eq:ContravariantSplitting}).

\subsection{Confirming the scale properties of the parameters}

We have obtained the expressions for the shape parameters, $\hat A,\hat B,\hat C$, and the scale parameters, $\mu_A$, in terms of the components of the contravariant emission metric. Now we can check that they satisfy the expected properties when some of the emitted time scales is rescaled.

Let us consider, for instance, that the first satellite does not emit the expected time, $\tau^1$, but emits a different time, related with the supposed one by
\[
	{\tau^1}'=f(\tau^1) \,.
\]
Then, the family of null cones emitted by this satellite will be relabeled and the natural covector will transform accordingly as
\[
	d{\tau^1}'=\dot{f} d\tau^1 \,,
\]
and the contravariant metric as
\[
	g^{1'A}=\dot{f} g^{1A} \,,
\]
with $g^{AB}$ unchanged for $A,B\neq 1$.

Taking the expression (\ref{eq:ShapeParameters}) of the shape parameters, for instance,
\[
	\hat A=\frac{2ABC}{\cal H} A \,,
\]
we can check its invariance:
\[
	{\cal H}'={\dot{f}}^2 {\cal H} \ ,\quad
	A'=\sqrt{\dot{f}} A \ ,\quad B'=\sqrt{\dot{f}} B \ ,\quad 
	C'=\sqrt{\dot{f}} C \quad
	\Rightarrow\quad \hat A'=\hat A \,.
\]
And taking the expression (\ref{eq:ScaleParameters}) of the scale parameters:
\[
	\mu^1=\sqrt{\frac{\cal H}{2g^{23}g^{24}g^{34}}}\quad
	\mbox{and}\quad
	\mu^2=\sqrt{\frac{\cal H}{2g^{13}g^{14}g^{34}}} \,,
\]
we can check that
\[
	\mu^{1'}=\dot{f}\,\mu^1 \quad\mbox{and}\quad \mu^{2'}=\mu^2 \,.
\]

\section{Conclusion}

In this paper we have studied the causal and geometric nature of the natural basis and cobasis of emission coordinates in four-dimensional space-times, generated by four emitters broadcasting their time. We have obtained some interesting non-trivial properties that characterize this class of coordinates. The most remarkable result is the obtention of a natural splitting of the metric which involves the separation of its 6 degrees of freedom into 2 different types of parameters, which behaves in clearly different ways when the particular time scales broadcasted by the emitters is changed: 2 parameters independent of the four time scales and four parameters, one for each emitter, dependent only on the corresponding scales. This parametrization should simplify the construction of the positioning systems, in decoupling those characteristics depending on the trajectories of the emitters from those that  depend on the time broadcasted. 

The relativistic positioning systems based in emission coordinates are quite well developed for two-dimensional space-times. However, the known results for the two-dimensional case are not trivially generalizable for higher dimensions.
This is the first quantitative article aimed to develop a theory of relativistic positioning systems in four-dimensional space-times. In a following paper we will present properties relating the perception of emission coordinates by an arbitrary observer and the existence of special observers \cite{CollPozo06d}. Also we will present some global properties of this coordinates implied by the condition of causality in the space-time \cite{CollPozo06c}, and will study the positioning system obtained for some simple particular cases in Minkowski space-times \cite{CollPozo06b}.

\ack
J.M.Pozo acknowledges the support of the postdoc fellowship EX-2004-0090, from the spanish {\it Ministerio de Educaci\'on y Ciencia}, and the project BFM2003-07076.

\end{document}